\pdfoutput=1

\documentclass[a4paper]{raa}

\usepackage[pass,paperwidth=8.27in,paperheight=11.69in]{geometry}
\usepackage{graphicx,times}
\usepackage{natbib}
\usepackage{amssymb,amsmath}
\usepackage{a4pdf}
\bibpunct{(}{)}{;}{a}{}{,}

\begin{document}

\title{Dark {halos} acting as chaos controllers in
asymmetric triaxial galaxy models}

 \volnopage{ {\bf 2011} Vol.\ {\bf 11} No. {\bf 7}, 811--823}
   \setcounter{page}{811}

\author{Nicolaos D. Caranicolas 
\and Euaggelos E. Zotos}

\institute{Department of Physics, Section of  Astrophysics,
Astronomy and Mechanics, Aristotle University of Thessaloniki 541
24, Thessaloniki, Greece;
{\it evzotos@astro.auth.gr}\\
\vs \no
   {\small Received 2011 January 2; accepted 2011 April 2}
}

\abstract{  We study the regular or chaotic character of orbits in
a 3D dynamical model, describing a triaxial galaxy surrounded by a
spherical dark halo component. Our numerical experiments suggest
that the percentage of chaotic orbits decreases exponentially as
the mass of the dark halo increases. A linear increase of the
percentage of the chaotic orbits was observed as the scale length
of the halo component increases. In order to distinguish between
regular and chaotic motion, we chose to use the total angular
momentum $L_{\rm tot}$ of the 3D orbits as a new indicator.
Comparison with other, previously used, dynamical indicators, such
as the Lyapunov Characteristic Exponent
 or the $P(f)$ spectral method, shows that the
$L_{\rm tot}$ indicator gives very fast and reliable results for
characteriz{ing} the nature of orbits in galactic dynamical
models.
  }
\authorrunning{N. D. Caranicolas \& E. E. Zotos}
\titlerunning{Dark Halos Acting as Chaos Controllers
in Asymmetric Triaxial Galaxy Models}
\maketitle

\section{Introduction}

In this paper we shall study the motion in a 3D composite galaxy
model described by the potential
\begin{equation}
{{V}_{t}}\left( x,y,z \right)={{V}_{g}}\left(x,y,z \right)
+{{V}_{h}}\left(x,y,z \right) ,
\end{equation}
where
\begin{equation}
{{V}_{g}}\left(x,y,z \right) =\frac{\upsilon _{0}^{2}}{2}\ln \left(
{{x}^{2}} -\lambda {{x}^{3}}+ \alpha {{y}^{2}}+b{{z}^{2}}+c_{b}^{2}
\right),
\end{equation}
while
\begin{equation}
{{V}_{h}}\left(x,y,z\right)
=\frac{-{{M}_{h}}}{{{\left( {{x}^{2}}+{{y}^{2}}+{{z}^{2}}+c_{h}^{2} \right)}^{1/2}}}.
\end{equation}
Potential {Equation~}(2) describes a triaxial elliptical galaxy
with a bulge and a small asymmetry introduced by the term
$-\lambda x^3$, $\lambda \ll 1$ (see \citealt{Binney2008}). The
parameters $\alpha${ and} $b$ describe the flattening of the
galaxy, while $c_b$ is the scale length of the bulge of the
galaxy. The parameter $\upsilon _0$ is used for the consistency of
the galactic units. To this potential we add a spherical dark
halo, described by the potential {Equation}~(3). Here $M_h$ and
$c_h$ are the mass and the scale length of the dark halo
component, respectively.

The aim of this article is twofold: (i) {T}o investigate the motion in
the potential {Equation}~(1) and to determine the role played by
the halo on the character of orbits. In particular, we are
interested {in} connect{ing} the percentage of
chaotic orbits, as well as the degree of chaos, with the
physical parameters, such as the mass and the scale length of the
dark halo component. (ii) To introduce, use and check a new fast
indicator, which is the total angular momentum $L_{\rm tot}$, of the
3D orbits, in order to obtain a reliable criterion to distinguish
between ordered and chaotic orbits.

The outcomes of the present research are mainly
based on the numerical integration of the equations of
motion
\begin{eqnarray}
\ddot{x}&=&-\frac{\partial \ V_t(x,y,z)}{\partial x}, \nonumber \\
\ddot{y}&=&-\frac{\partial \ V_t(x,y,z)}{\partial y},  \\
\ddot{z}&=&-\frac{\partial \ V_t(x,y,z)}{\partial z}\nonumber ,
\end{eqnarray}
where the dot indicates derivatives with respect to time. The
Hamiltonian {of} the potential {Equation}~(1){ is}
{written as}
\begin{equation}
H=\frac{1}{2}\left(p_x^2+p_y^2+p_z^2\right)+V_t(x,y,z)=h_3,
\end{equation}
where $p_x$, $p_y$ and $p_z$ are the momenta per unit mass
conjugate to $x$, $y$ and $z${,} respectively, while $h_3$ is the
numerical value of the Hamiltonian.

In this article, we use a system of galactic units, where the unit
of length is $1$\,kpc, the unit of mass is $2.325 \times 10^7
M_{\odot}$ and the unit of time is $0.97748 \times 10^8$\,yr.
The velocity unit is $10$\,km~s$^{-1}$, while $G$ is equal to
unity. The energy unit (per unit mass) is
$100$\,{km$^{2}$s$^{-2}$}. In the above units
we use the values: $\upsilon _0=15$, $c_b  =2.5$, $\alpha =1.5$,
$b=1.8${ and}  $\lambda =0.03$, while $M_h${ and} $c_h$ are treated as
parameters. Orbit calculations are based on the numerical
integration of the equations of motion{ given in} {Equation}~(4). This
was made using a Bulirsh - Stoer method in double precision and
the accuracy of the calculations was checked by the constancy of
the energy integral {Equation}~(5), which was conserved up to
the twelfth significant figure.

The article is organized as follows: In Section~2 we introduce a new
dynamical parameter{ and} present
results for the 2D system. In Section~3 we study the character of
orbits in the 3D system{ and make a} comparison with
other indicators. Finally, in
Section~4, we present a discussion and the conclusions of this
research.

\section{A new indicator: The character of motion in the 2D system}

The value of the total angular momentum for a star of mass $m=1$
moving in a 3D orbit is
\begin{equation}
L_{\rm tot}=\sqrt{L_x^2 + L_y^2 + L_z^2} ,
\end{equation}
where $L_x$, $L_y$,{ and} $L_z$ are the three components of angular
momentum along the $x$, $y$ and $z$ ax{e}s{, respectively,} given by
\begin{eqnarray}
L_x &=& y\dot{z} - \dot{y}z, \nonumber \\
L_y &=& z\dot{x} - \dot{z}x, \\
L_z &=& x\dot{y} - \dot{x}y\nonumber \, .
\end{eqnarray}
For {the} 2D system we {describe} in {Equation}~(7),
$z=\dot{z}=0$, that is $L_{\rm tot}$ reduces to $L_z$. Here, we
must note that the total angular momentum {Equation}~(6) is
conserved only for a spherical system. The same is true for all
the three components of the angular momentum. On the other hand,
in axially symmetric galactic models only the $L_z$ component of
the angular momentum is conserved. In this research, we shall use
the plot of the $L_{\rm tot}$ vs. time in order to distinguish
regular from chaotic motion.

Our next step is to study the properties of the 2D dynamical
system, which comes from potential {Equation}~(1) if we
set $z=0$. The corresponding 2D Hamiltonian {is
written as}
\begin{equation}
H_2=\frac{1}{2}\left(p_x^2+p_y^2\right)+V_t(x,y)=h_2 ,
\end{equation}
where $h_2$ is the numerical value of the Hamiltonian. We do this in
order to use the results obtained for the 2D model in the study of
the more complicated 3D model, which will be presented in the next
{s}ection.

\begin{figure}

\vs\vs \centering

\includegraphics[width=70mm]{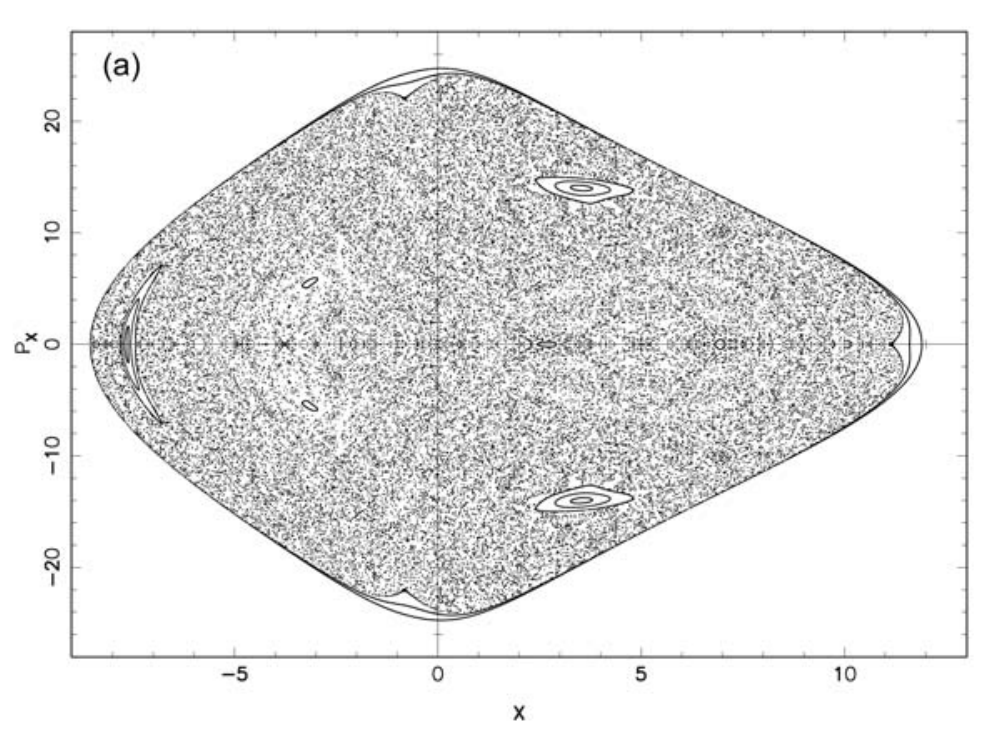}~~~
\includegraphics[width=70mm]{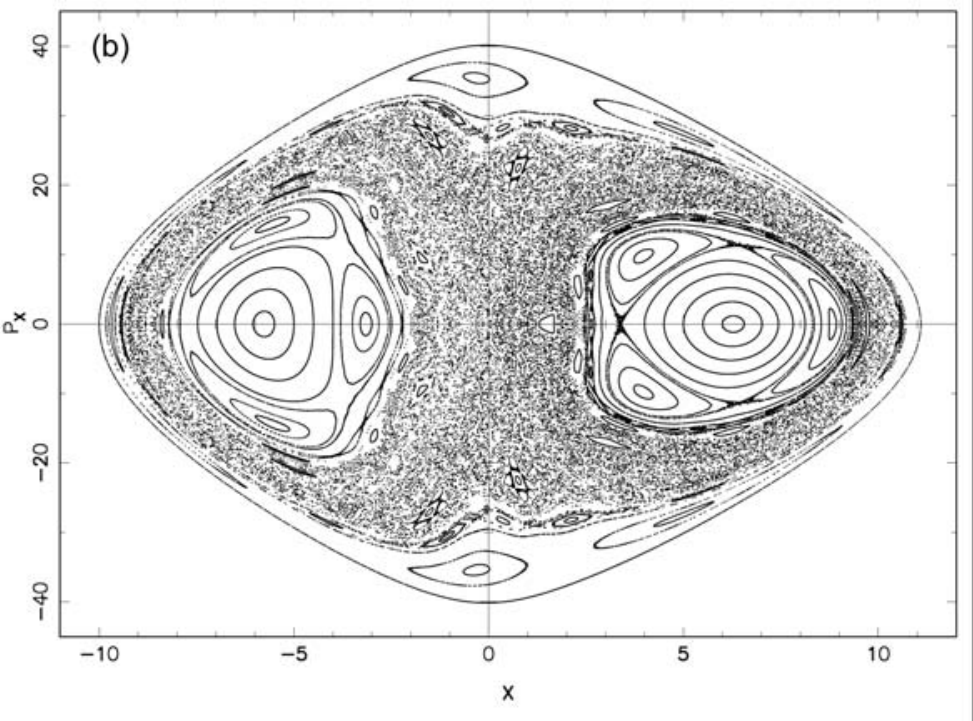}

\includegraphics[width=70mm]{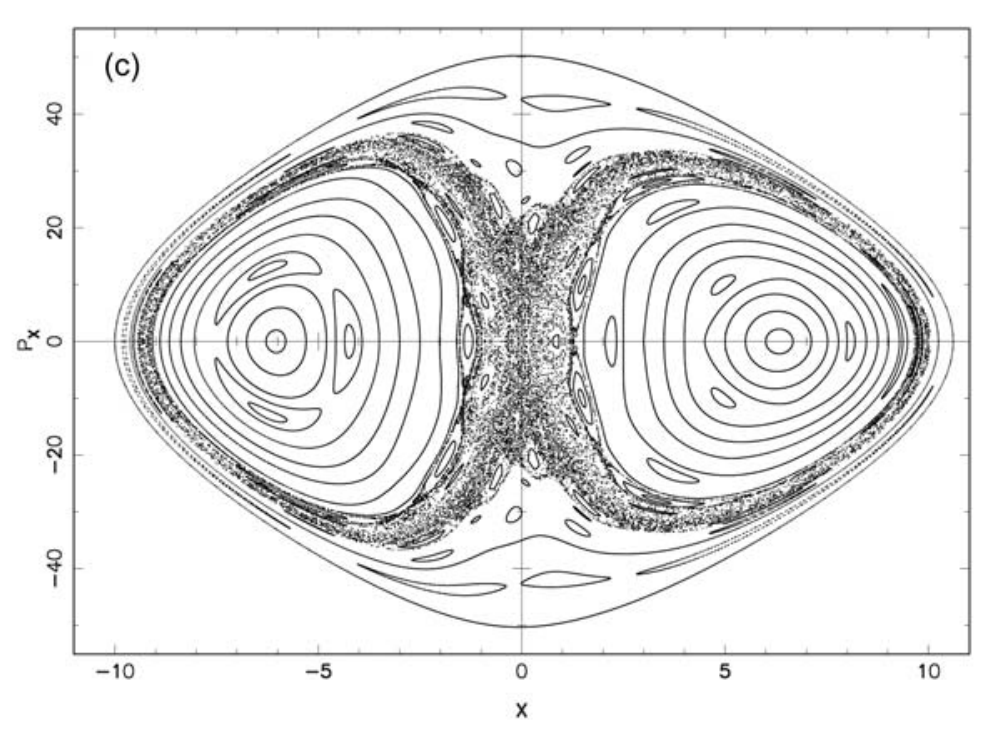}~~~
\includegraphics[width=70mm]{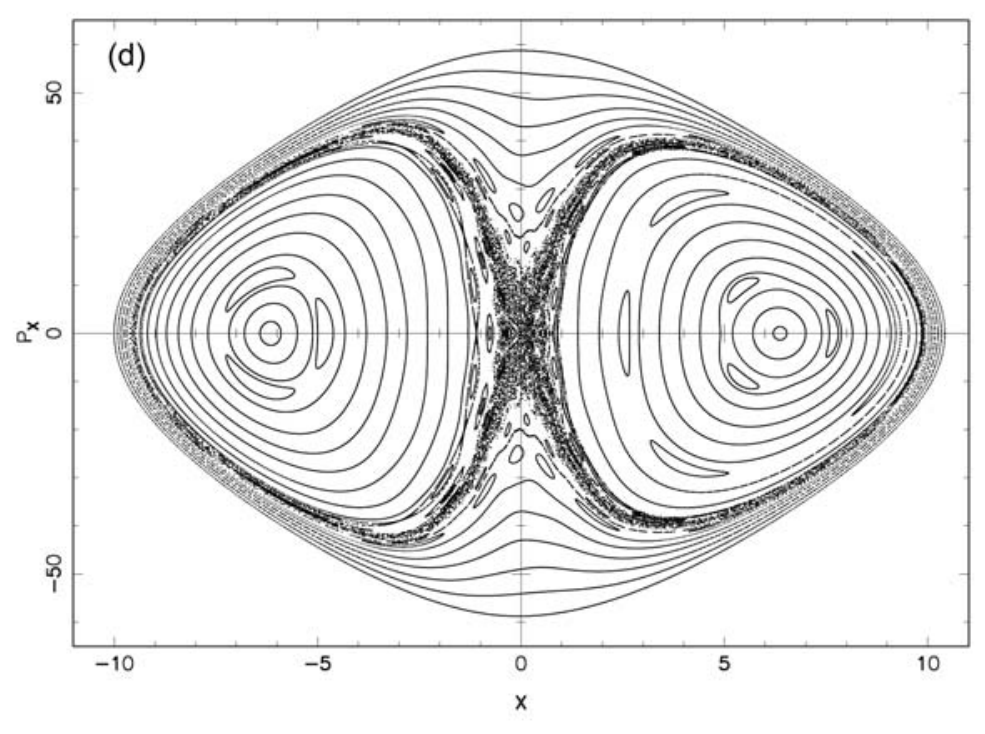}


\caption{ \baselineskip 3.6mm The $(x,p_x)$ phase plane when (a) 
$M_h=0$, $h_2=516$; (b) 
 $M_h=10\,000$, $h_2=-226$; (c) 
 $M_h=20\,000$, $h_2=-1007$ and
(d) 
$M_h=30\,000$, $h_2=-1788$. The values of all other parameters are
given in {the }text. }
\end{figure}

\begin{figure}

\vs\vs

\centering
\includegraphics[width=70mm]{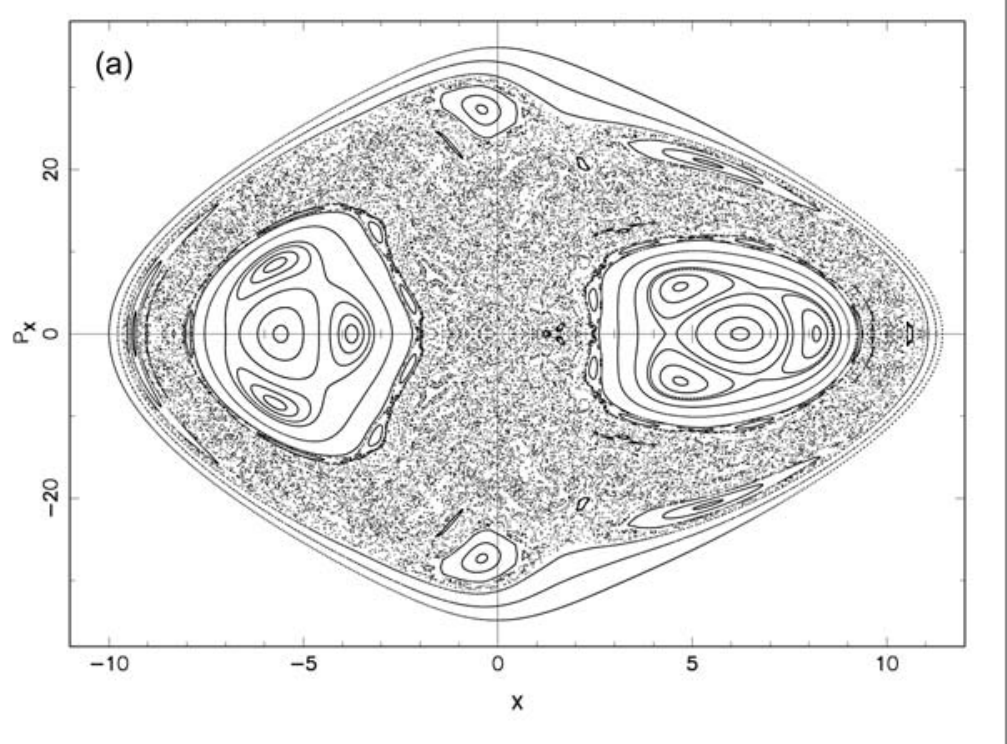}~~~
\includegraphics[width=70mm]{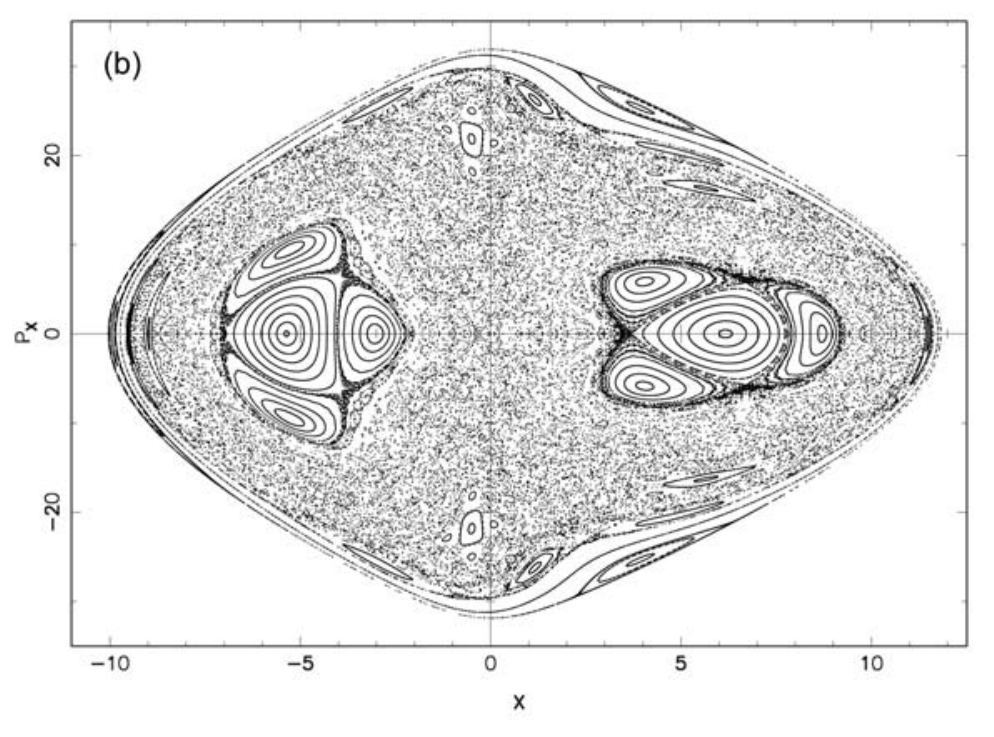}

\includegraphics[width=70mm]{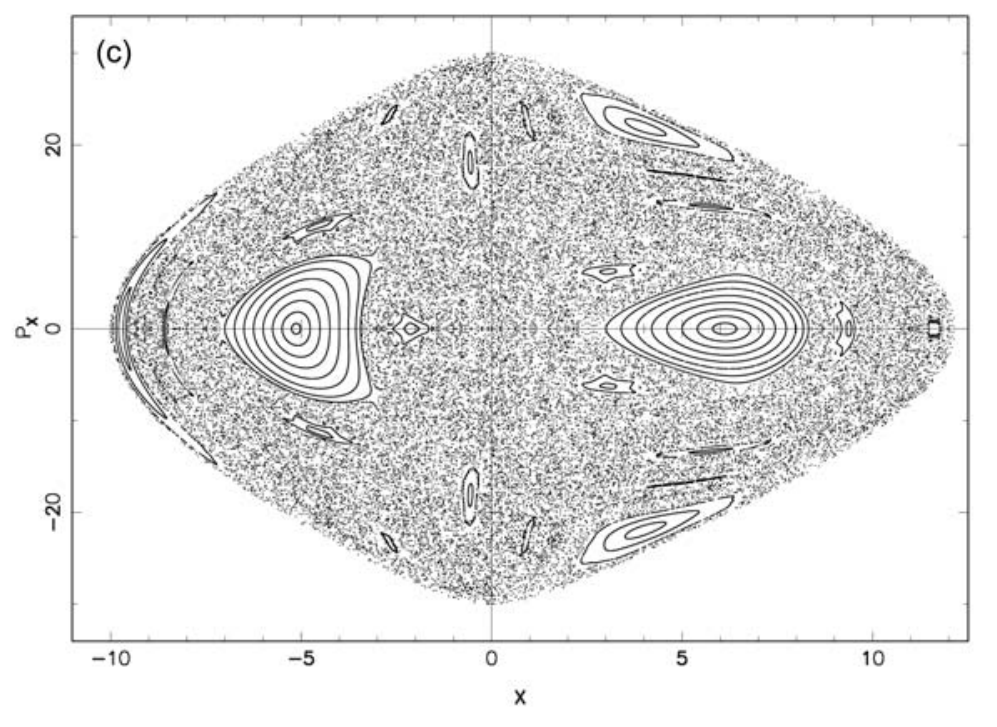}~~~
\includegraphics[width=70mm]{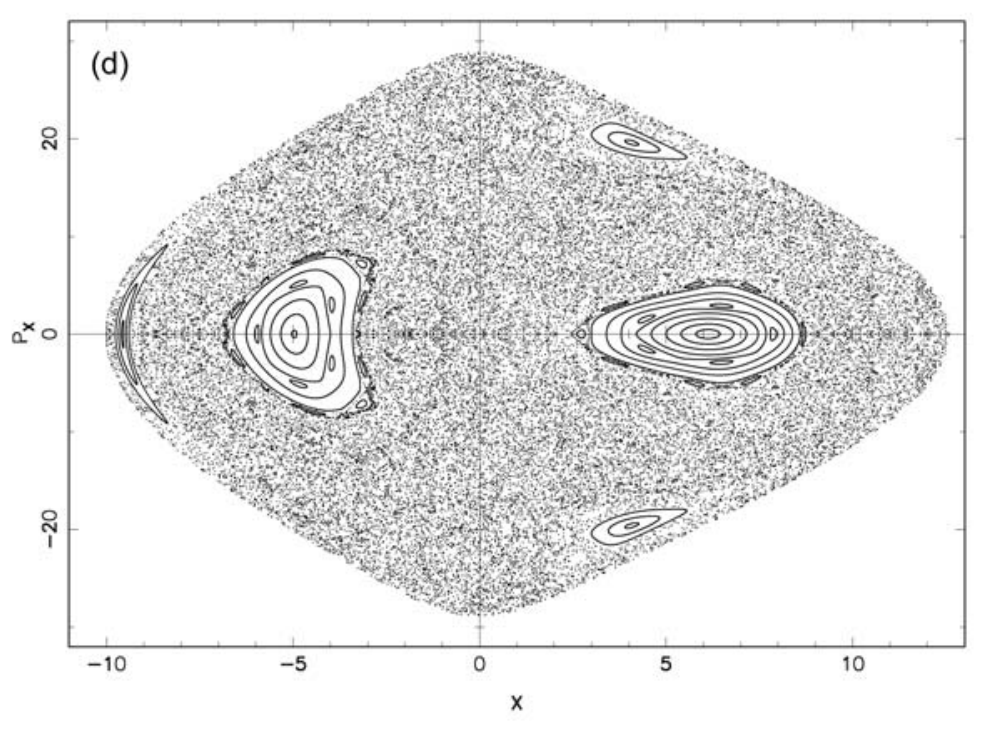}


\caption{\baselineskip 3.6mm The $(x,p_x)$ phase plane, when
$M_h=10\,000$ and (a) 
$c_h=10.5$, $h_2=-135$; (b) $c_h=13$, $h_2=-55$; (c)
$c_h=15.5$, $h_2=11$ and (d) 
$c_h=18$, $h_2=68$. The values of all other parameters are given
in the text.}
\end{figure}

Figure~1(a)--(d) 
   shows the $(x,p_x)$,
$\left(y=0,p_y>0\right)$ phase plane for four different values of
the mass of the halo. The values of all other parameters are
$\upsilon _0=15$, $c_b=2.5$, $\alpha =1.5$, $b=1.8$,
$\lambda=0.03$,{ and} $c_h=8$. The values of the energy $h_2$ were
chosen so that in all phase planes $x_{\rm max} \simeq 10$.
Figure~1(a) shows the phase plane when the system has no halo
component, that is when $M_h=0$. The value of $h_2$ is 516. One
can see that almost all {of }the phase plane is covered by a
chaotic sea. The regular regions consist of {a} small set of
islands produced by secondary resonances. Figure~1(b) shows the
phase plane when $M_h=10\,000$, while $h_2=- 226$. As we{ can}
see, the majority of the phase plane is covered by chaotic orbits.
There are also two considerable regular regions inside the chaotic
sea. These  belong  to  invariant curves produced by
quasi-periodic orbits which are characteristic of{ a} 1:1
resonance. There are also regular regions produced by
quasi-periodic orbits{ which are} characteristic of {a} 2:3
resonance. In the outer part we can see some regular regions
produced by quasi-periodic orbits characteristic of the 2:1
resonance. Some small islands produced by secondary resonances are
also embedded in the chaotic sea. Figure~1(c) is similar to
Figure~1(a) but when $M_h=20\,000$ and $h_2=-1007$. Here, the
chaotic region is much smaller, while the majority of orbits are
regular. The most prominent characteristic of this phase plane is
the presence of many small islands produced by secondary
resonances. Figure~1(d) shows the phase plane when $M_h=30\,000${
and} $h_2=-1788$. Here we{ only} see a small chaotic layer, while
the rest of the phase plane is covered by regular orbits. The
characteristic of this phase plane is that a considerable part of
the regular orbits are box orbits. Secondary resonances are also
observed.

Therefore, our numerical results suggest that the chaotic regions
in our 2D composite galactic dynamical system described by the
Hamiltonian {Equation~}(8) strongly depend on the mass
of the halo component. The mass of the halo acts as a catalyst on
the asymmetry of the galaxy and drastically reduces the percentage
of the chaotic orbits. Thus, one can conclude that massive
spherical dark halos can act as chaos
controllers in galaxies showing small asymmetries.

\begin{figure*}
\centering


\includegraphics[width=68mm]{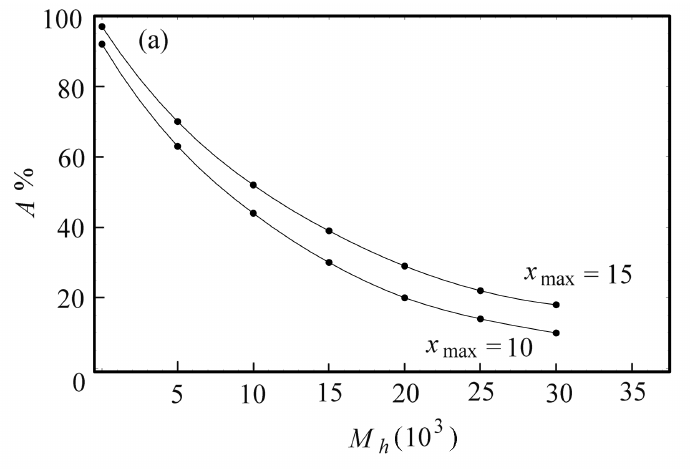}~~
\includegraphics[width=68mm]{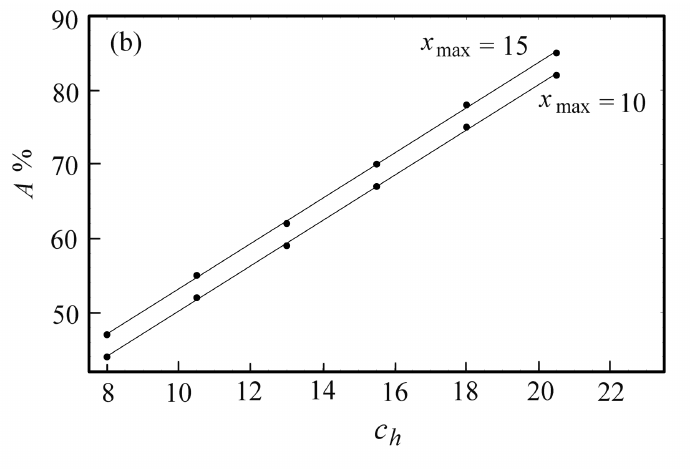}

\caption{\baselineskip 3.6mm (a) 
A plot of the area $A\%$ covered by chaotic orbits vs. $M_h$ and
(b) 
A plot of the area $A\%$ covered by chaotic orbits vs. $c_h$. The
values of all {the }other parameters are given in text.}
\end{figure*}

Figure~2(a)--(d)   is similar to Figure~1(a)--(d), when $M_h$ is
10\,000 while $c_h$ is treated as a parameter. All other
parameters are as{ shown} in Figure~1. Here again, the values of
the energy $h_2$ were chosen so that in all phase planes $x_{\rm
max} \simeq 10$. In Figure~2(a) we have $c_h=10.5${ and}
$h_2=-135$. Here the phase plane has a large chaotic region, while
one also observes considerable areas of regular motion. In
Figure~2(b), where the values of $c_h$ and $h_2$ are 13 and
--55{,} respectively, the chaotic sea increases, while the regular
region decreases. In the phase plane shown in Figure~2(c), we have
taken $c_h=15.5${ and} $h_2=11$. Obviously, the chaotic sea is
larger than that shown in Figure~2(b). On the other hand, the
regular region is smaller than that given in Figure~2(b). Finally,
in the results presented in Figure~2(d) we have chosen $c_h=18${
and} $h_2=68$. Here the chaotic sea is even larger, while the
regular regions are smaller than those shown in Figure~2(c).

The conclusion is that, for a given mass of the dark halo
component, the percentage of chaotic orbits increases as the scale
length of the halo increases. In other words, the numerical
experiments indicate that one would expect to observe less
chaos in asymmetric triaxial galaxies surrounded by dense halos,
while the chaotic orbits would increase in similar galaxies
surrounded {by} less dense spherical halo components.

Figure~3(a) 
  shows the percentage of the phase
plane $A\%$ covered by chaotic orbits as a function of the mass of
the dark halo, for two different values of $x_{\max}$. The values
of the parameters are $\upsilon_0 = 15$, $c_b = 2.5$, $\alpha =
1.5$, $b = 1.8$, $\lambda = 0.03$  and $c_h = 8$. We see that
$A\%$ decreases exponentially as $M_h$ increases. Figure~3(b)
shows a plot {of} $A\%$ and $c_h$. The values of the parameters
are $\upsilon_0 = 15$, $c_b = 2.5$, $\alpha = 1.5$, $b = 1.8$,
$\lambda = 0.03$  and $M_h = 10\,000$. Here we see that $A\%$
increases linearly as $c_h$ increases. The authors would like to
make {it }clear that $A\%$ is estimated on a completely empirical
basis by measuring the area in the $(x,p_x)$ phase plane occupied by
chaotic orbits.

Figure~4(a) and (b) show{s} a plot of the Lyapunov Characteristic
Exponent (LCE) (see \citealt{Lichtenberg1992}) vs. $M_h$ {and} 
$c_h${,} respectively. The values of the parameters for
Figure~4(a) {are the same }as in Figure~3(a) and for Figure~4(b)
are {the same }as in Figure~3(b). One can see, in Figure~4(a),
that the LCE decreases exponentially {when} $M_h$ increases, while
in Figure~4(b) we see that the LCE increases exponentially {when}
$c_h$ increases. Here we must point out that it is well known that
the LCE has different values in each chaotic component (see
\citealt{Saito1979}). {Since} we have regular regions{ in all
cases} and only a large chaotic sea, we calculate the average
value of the LCE in each case by taking thirty orbits with
different initial conditions in the chaotic sea. In all cases, the
calculated values {of} the LCEs were different in the fourth
decimal point in the same chaotic sea.

\begin{figure*}
\centering
\includegraphics[width=68mm]{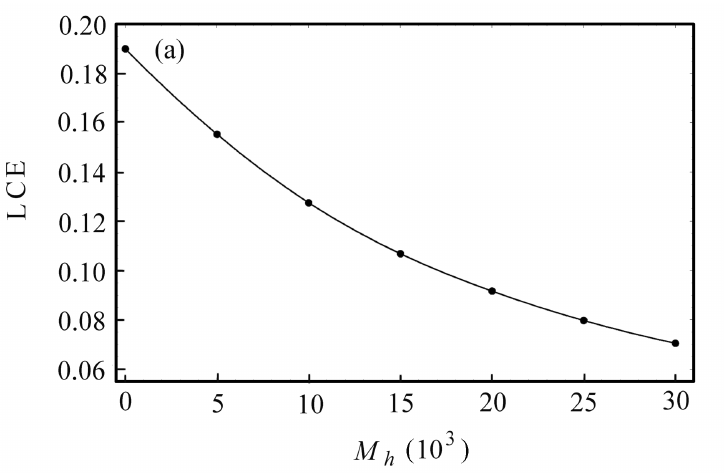}
\includegraphics[width=68mm]{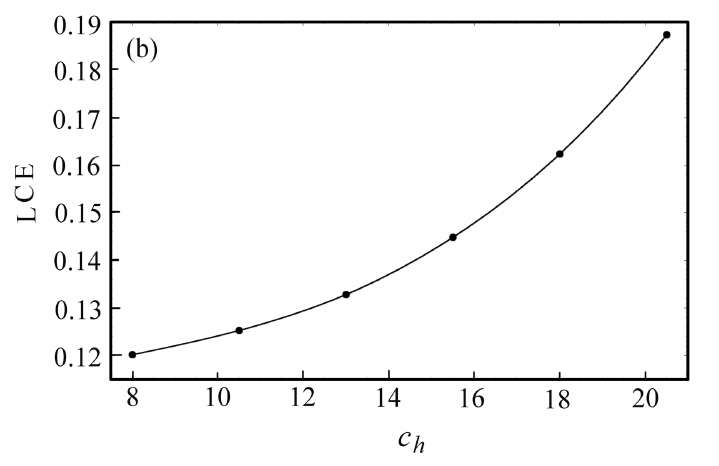}

\vspace{-3mm}

\caption{\baselineskip 3.6mm (a) 
A plot of the LCE vs. $M_h$ and (b) 
A plot of the LCE vs. $c_h$. The values of all {the }other
parameters are given in the text.}

\vs
 \centering
\includegraphics[width=66mm]{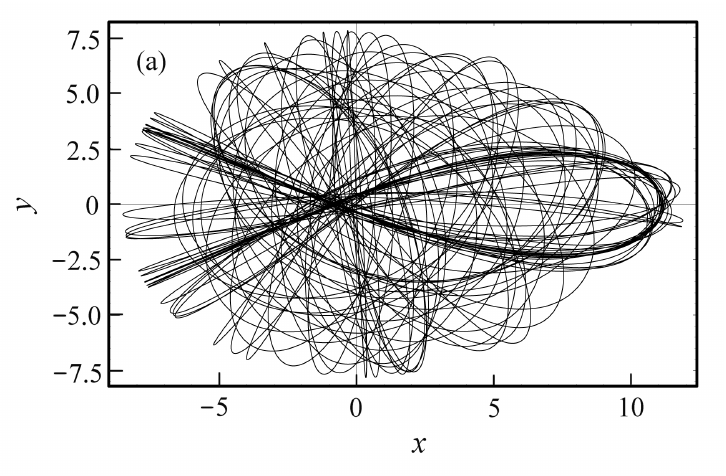}
\includegraphics[width=66mm]{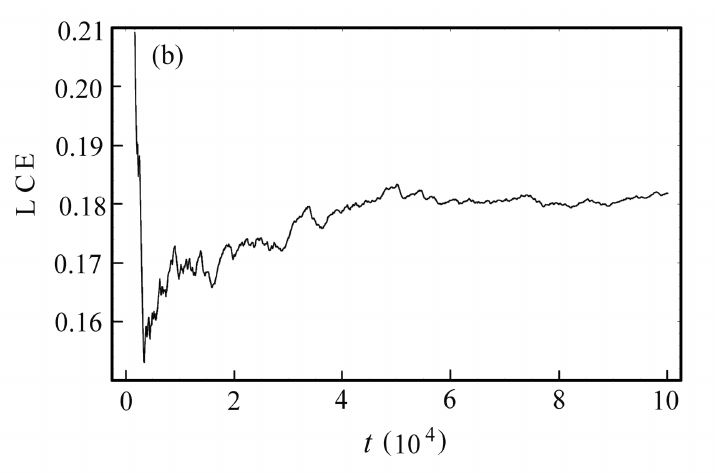}

\includegraphics[width=66mm]{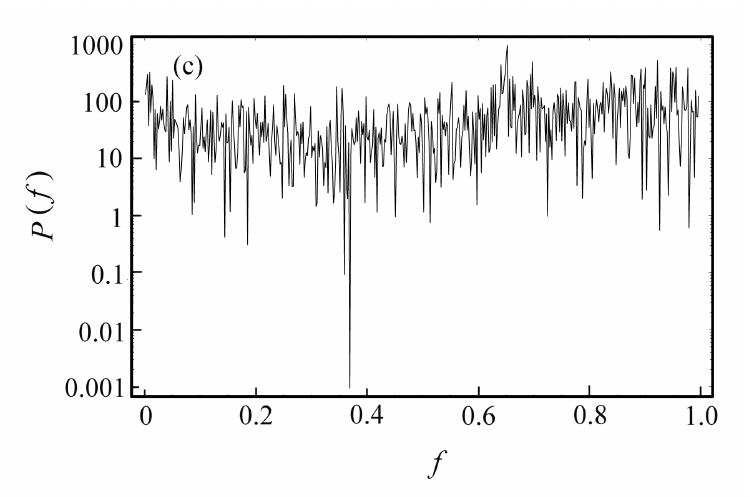}
\includegraphics[width=66mm]{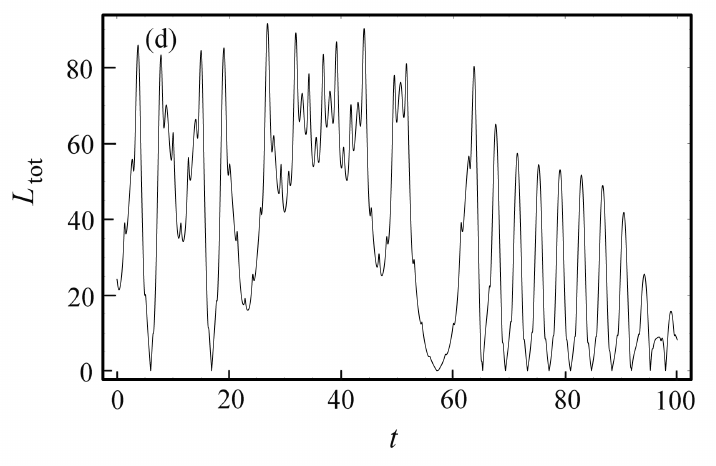}

\vspace{-3mm}

\caption{\baselineskip 3.6mm (a) 
An orbit in the 2D potential, (b) 
Corresponding LCE, (c) 
$P(f)$ indicator and (d) 
$L_{\rm tot}$ indicator. The motion {is} chaotic. See text for
details.}
\end{figure*}

In the following, we shall investigate the regular or chaotic
character of orbits in the 2D Hamiltonian {Equation~}(8) using the
new dynamical indicator $L_{\rm tot}$. In order to see the
effectiveness of the new method, we shall compare the results with
two other indicators, the classical method of the LCE and the
method $P(f)$ used by \cite{Karanis2007}. This method uses the
Fast Fourier Transform (F.F.T{.}) of a series of time intervals,
each one representing the time that elapsed between two successive
points on the Poincar\`{e} $(x,p_x)$ phase plane for 2D systems,
while for 3D systems they take two successive points on the plane
$z=0$.

Figure~5(a) shows an orbit with initial conditions $x_0=-1.0${
and} $y_0=p_{x0}=0$, while the value of $p_y$ is always found from
the energy integral for all orbits. The values of all other
parameters and energy are the same as in Figure~1(a). One observes
in Figure~5(b) that the LCE, which was computed for a period of
$10^5$ time units, has a value of about 0.18{,} indicating chaotic
motion. The same result is shown by the $P(f)$ indicator, which is
given in Figure~5(c). Figure~5(d) shows a plot of the $L_{\rm
tot}$ vs. time for a time interval of 100 time units. We see that
the diagram is highly asymmetric. Furthermore, one observes large
deviations between the maxima and also large deviations between
the minima in the $[L_{\rm tot}, t]$ plot. The above
characteristics suggest that the corresponding orbit is chaotic.

\begin{figure}

\vs \centering

\includegraphics[width=66mm]{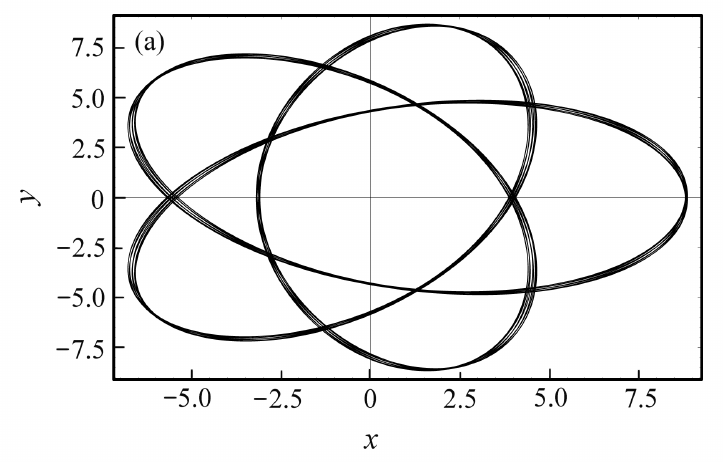}
\includegraphics[width=66mm]{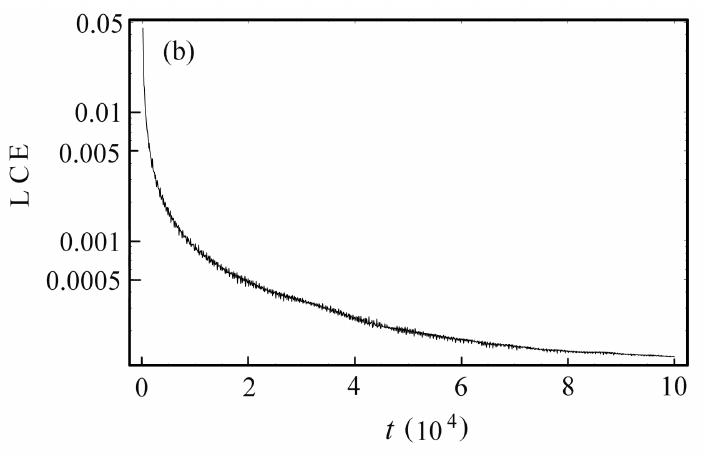}

\includegraphics[width=66mm]{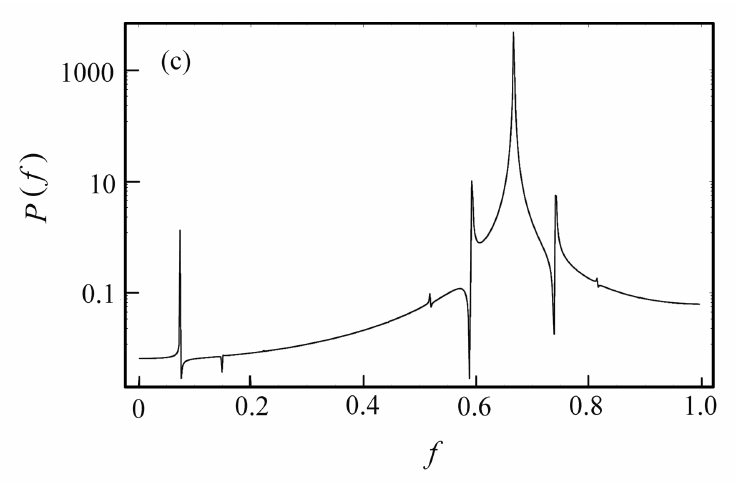}
\includegraphics[width=66mm]{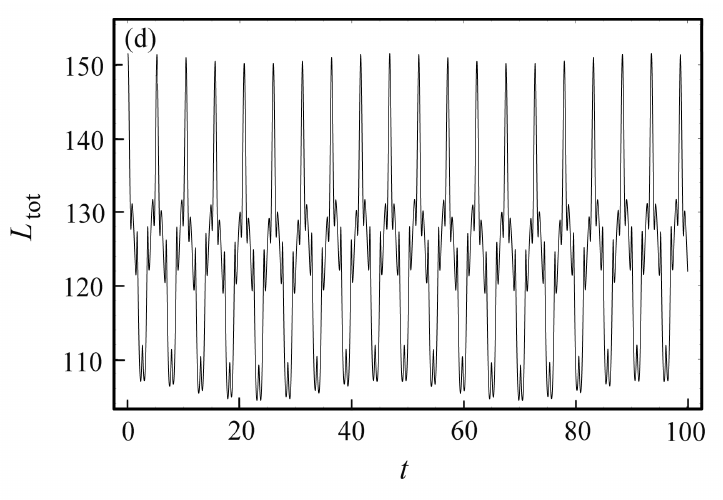}

\begin{minipage}[]{75mm}

\caption{ Similar {to} Fig.~5(a)--(d). The motion is
regular.}\end{minipage}

\end{figure}

Figure~6(a) shows an orbit with initial conditions $x_0=8.8$  and
$y_0=p_{x0}=0$. The values of all other parameters and energy are
{the same }as in Figure~1(b). As we{ can} see, this is a
quasi-periodic orbit. Therefore the LCE of this orbit goes to
zero, as is clearly seen in Figure~6(b). The $P(f)$ indicator in
Figure~6(c) shows a small number of peaks, also indicating regular
motion. The plot of the $L_{\rm tot}$ given in Figure~6(d) is now
quasi-periodic, with symmetric peaks, indicating regular motion.

\begin{figure}[h!!]

\centering

\includegraphics[width=67mm]{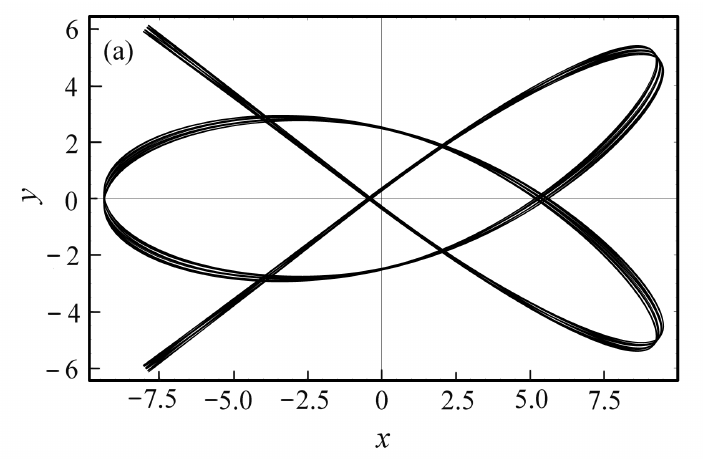}~~~~
\includegraphics[width=67mm]{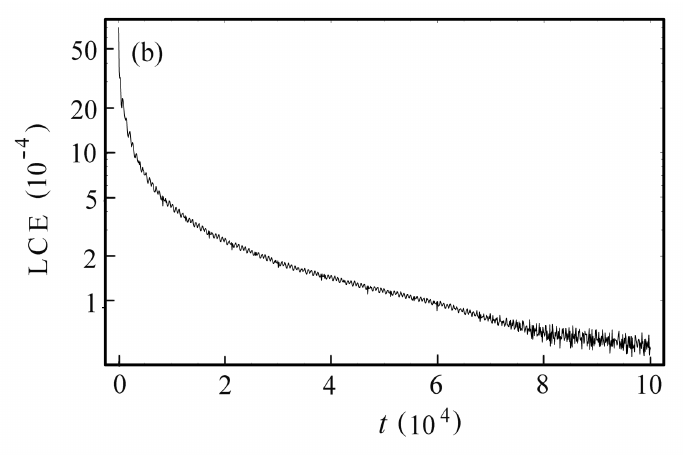}

\includegraphics[width=67mm]{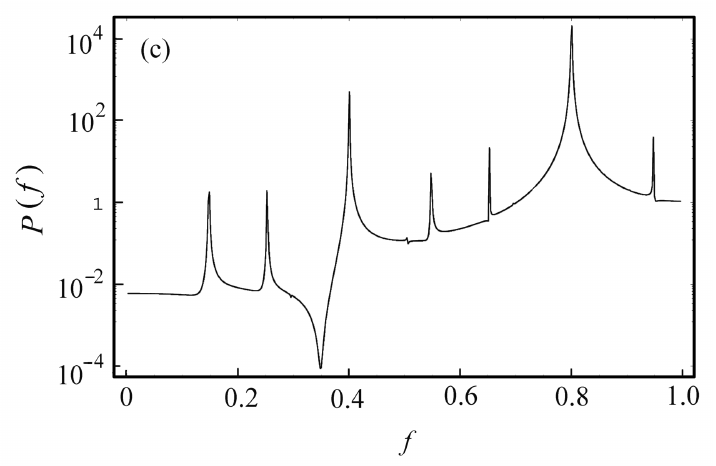}~~~~
\includegraphics[width=67mm]{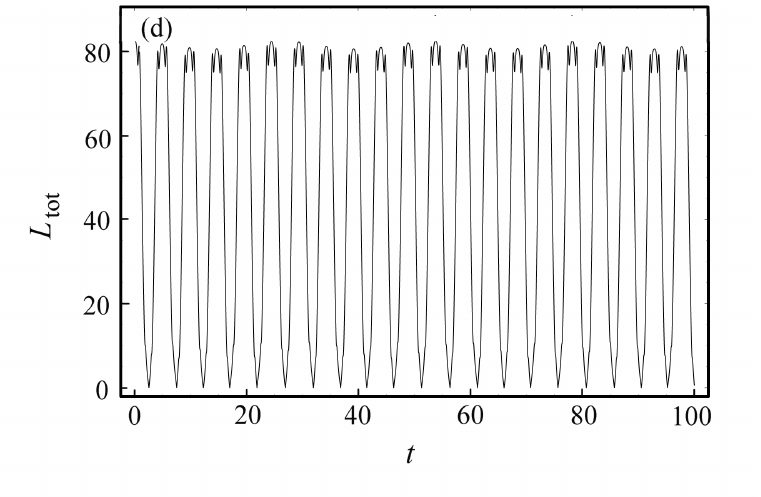}

\begin{minipage}[]{75mm}
\caption{ Similar {to} Fig.~6(a)--(d) for a regular
orbit.}\end{minipage}

\vs\vs \centering

\includegraphics[width=67mm]{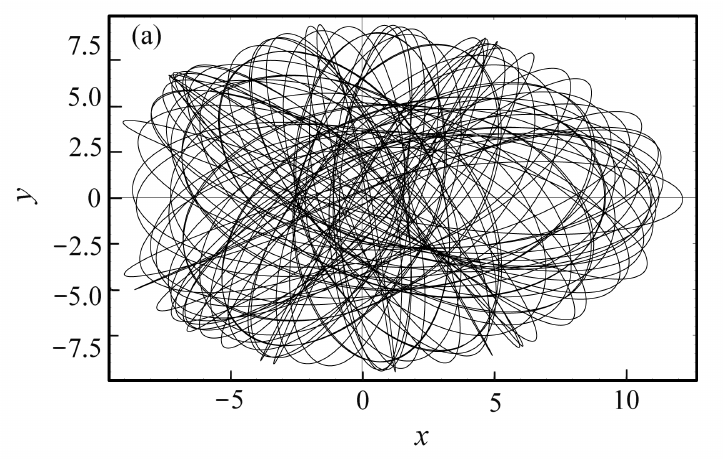}~~~~
\includegraphics[width=67mm]{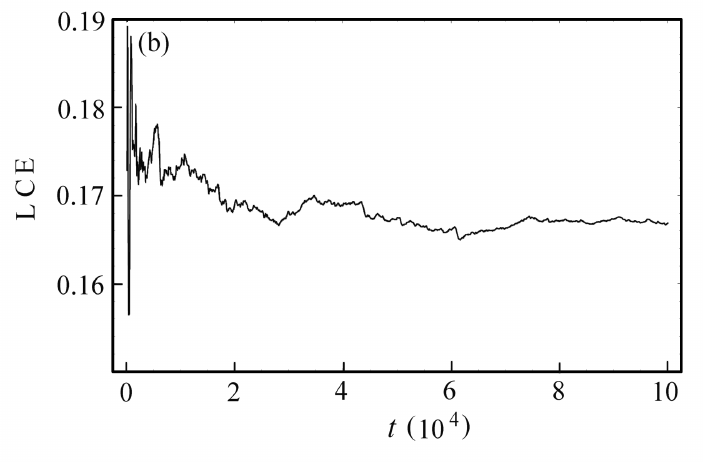}

\includegraphics[width=67mm]{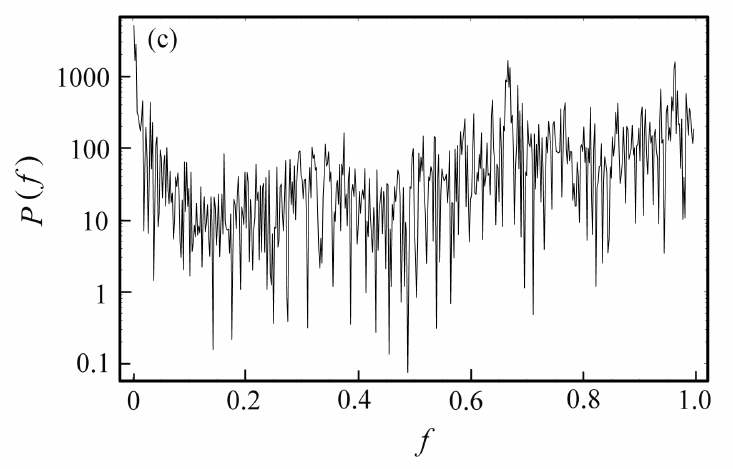}~~~~
\includegraphics[width=67mm]{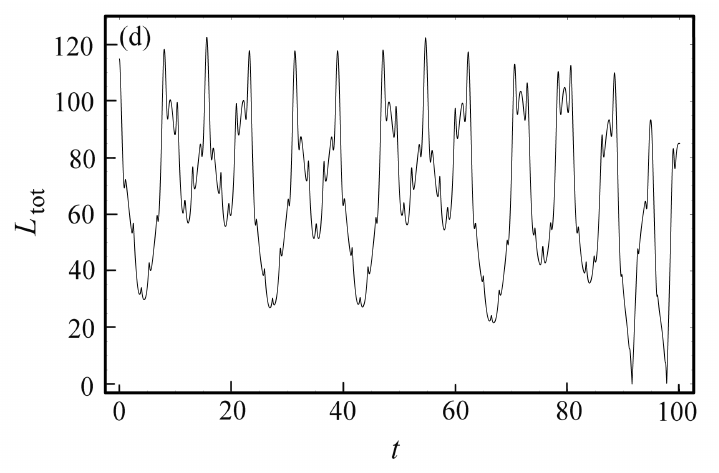}

\begin{minipage}[]{70mm}

\caption{ Similar {to} Fig.~5(a)--(d) for a chaotic
orbit.}\end{minipage}\vs
\end{figure}

Figure~7(a)--(d) is similar to Figure~6(a)--(d) for an orbit with
initial conditions $x_0=-9.36${ and} $y_0=p_{x0}=0$, while the
values of all other parameters and energy are {the same }as in
Figure~2(a). As we{ can} see, the orbit is quasi-periodic and this
fact is indicated by all three dynamical parameters. On the
contrary, the orbit shown in Figure~8(a) has initial conditions
$x_0=10${ and} $y_0=p_{x0}=0$ {while} the values of all other
parameters and energy are {the same }as in Figure~2(d). The orbit
looks chaotic and this is indicated by the LCE, the $P(f)$ and the
$L_{\rm tot}$ shown in Figure~8(b), 8(c) and 8(d), respectively.

A large number of orbits in the 2D system were calculated for
different values of the parameters. All {of the} numerical results
suggest that the  $L_{\rm tot}$ is a fast and reliable dynamical
parameter and can be safely used in order to distinguish ordered
from chaotic motion. There is no doubt that the $L_{\rm tot}$
method is much faster than the LCE, because it needs only about a
hundred time units, while the LCE needs about a hundred thousand
time units. {T}he $P(f)$ indicator needs several thousand time
units in order to give reliable results. Furthermore, it needs the
computation of the phase plane of the system, while the $L_{\rm
tot}$ {only }needs the calculations of the corresponding orbit.

\section{The character of motion in the 3D model}

We shall now proceed to investigate the regular or chaotic
behavior of the orbits in the 3D Hamiltonian {Equation}~(5).
The regular or chaotic nature of the 3D orbits is found as
follows: {W}e choose initial conditions
$\left(x_0,p_{x0},z_0\right)$, $y_0=p_{z0}=0$, such {that}
$\left(x_0,p_{x0}\right)$ is a point on the phase plane of the 2D
system. The point $\left(x_0,p_{x0}\right)$ lies inside the
limiting curve
\begin{equation}
\frac{1}{2}p_x^2+V_t(x)=h_2,
\end{equation}
which is the curve containing all the invariant curves of the 2D
system. We choose $h_3=h_2$ and the value of $p_{y0}$, for all
orbits, is obtained from the energy integral {Equation}~(5). Our
numerical experiments show that orbits with initial conditions
$\left(x_0,p_{x0},z_0\right)$, $y_0=p_{z0}=0$, such as
$\left(x_0,p_{x0}\right)$ which is a point in the chaotic regions
of Figures~1(a)--(d) and 2(a)--(d) for all permissible values of
$z_0$, produce chaotic orbits.

Our next step is to study the character of orbits with initial
conditions $\left(x_0,p_{x0},z_0\right),y_0=p_{z0}=0$, such as
$\left(x_0,p_{x0}\right)${ which} is a point in the regular
regions of Figure~1(a)--(d) and  Figure~2(a)--(d). It was found
that in all cases the regular or chaotic character of the above 3D
orbits depends strongly on the initial value $z_0$. Orbits with
small{ values} of $z_0$ are regular, while for large values of
$z_0$ they change their character and become chaotic. The general
conclusion, which is based on the results derived from a large
number of orbits, is that orbits with values of $z_0 \geq 0.75$
are chaotic, while orbits with values of $z_0<0.75$ are regular.

Figure~9(a) shows the LCE of the 3D system, as a function of the
mass of{ the} halo, for a large number of chaotic orbits when
$c_h=8$. The values of the parameters are {shown} in Figure~4(a).
We see that the LCE decreases exponentially as the mass of the
halo increases. Figure~9(b) shows {the} LCE as a function of $c_h$
when $M_h=10\,000$. The values of the parameters are {shown} in
Figure~4(b). Here the LCE increases exponentially as $c_h$
increases. The above results suggest that the degree of chaos
decreases in asymmetric triaxial galaxies with more dense and more
massive halo components.

Figure~10(a)--(d) is similar to Figure~8(a)--(d) but for a 3D
orbit. The orbit shown in Figure~10(a) looks chaotic. Initial
conditions are $x_0=2.0$, $p_{x0}=0${ and} $z_0=0.5$. Remember
that all orbits have $y_0=p_{z0}=0$, while the value of $p_{y0}$
is always found from the energy integral. The values of all other
parameters and energy $h_3$ are {the same }as in Figure~2(b). The
LCE shown in Figure~10(b) assures the chaotic character of the
orbit. The $P(f)$ given in Figure~10(c) also suggests chaotic
motion. The same conclusion comes from the $L_{\rm tot}$, which is
shown in Figure~10(d). Figure~11(a)--(d) is similar to
Figure~10(a)--(d) but for a quasi-periodic 3D orbit. The initial
conditions are $x_0=5.0$, $p_{x0}=0${ and} $z_0=0.1$. The values
of all other parameters and energy $h_3$ are {the same }as in
Figure~1(c). Here one observes that all three detectors support
the regular character of orbit{s}.

\begin{figure}[h!!!]

\vs \centering

\includegraphics[width=66mm]{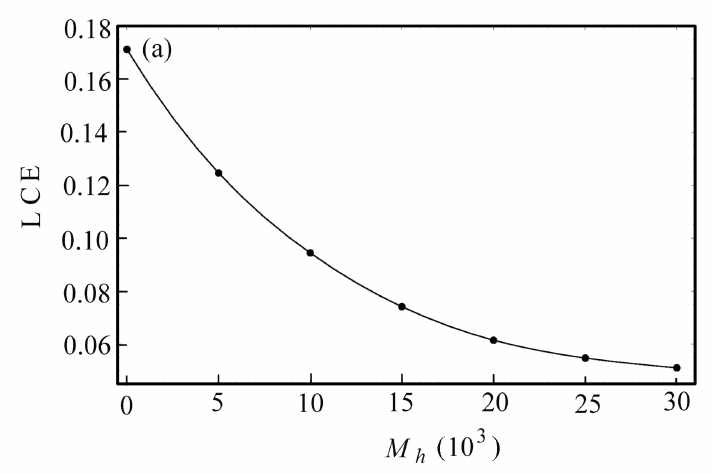}~~~~
\includegraphics[width=66mm]{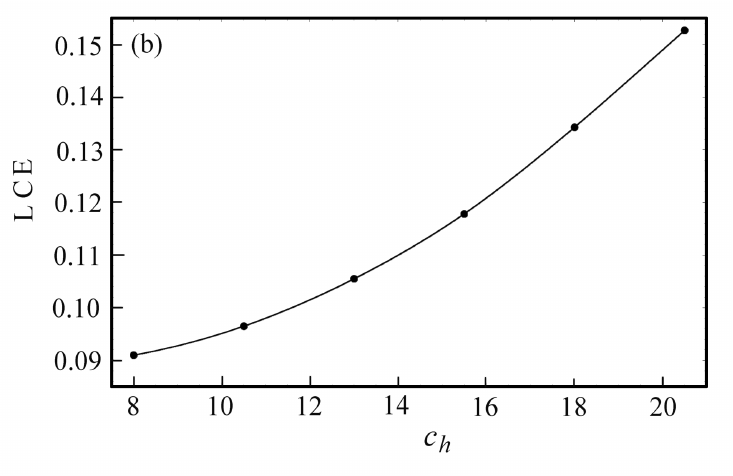}

\vs
\begin{minipage}[]{130mm}

\caption{\baselineskip 3.6mm
 Similar to Fig.~4(a)--(b) for the 3D
potential. The values of all other parameters are given in the
 text.}\end{minipage}

\vs\vs
 \centering

\vs \vs
\includegraphics[width=58mm]{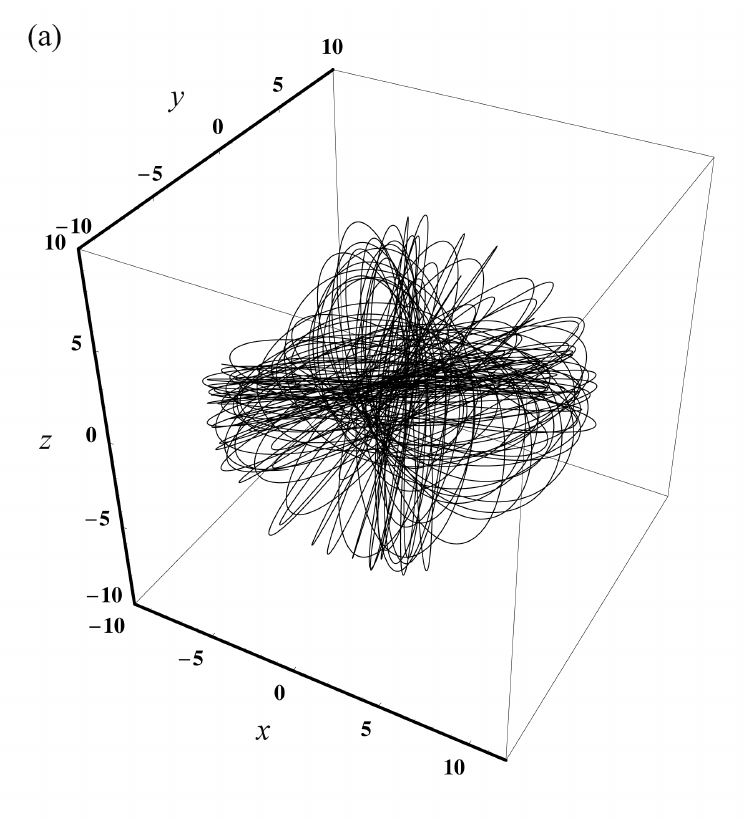}
\includegraphics[width=68mm]{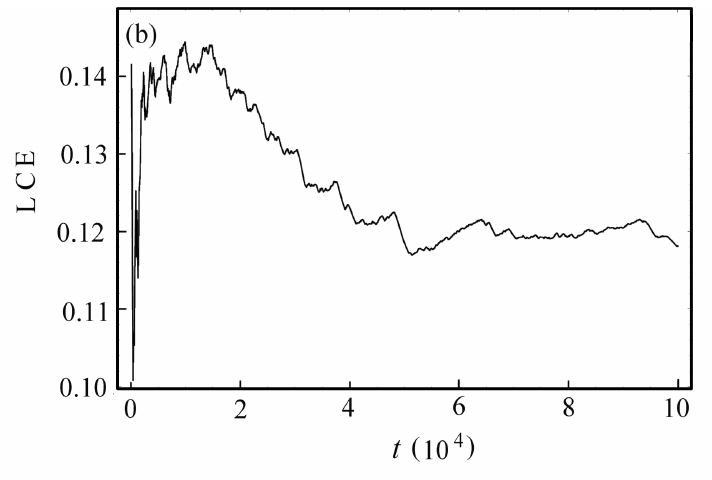}

\vs

\includegraphics[width=66mm]{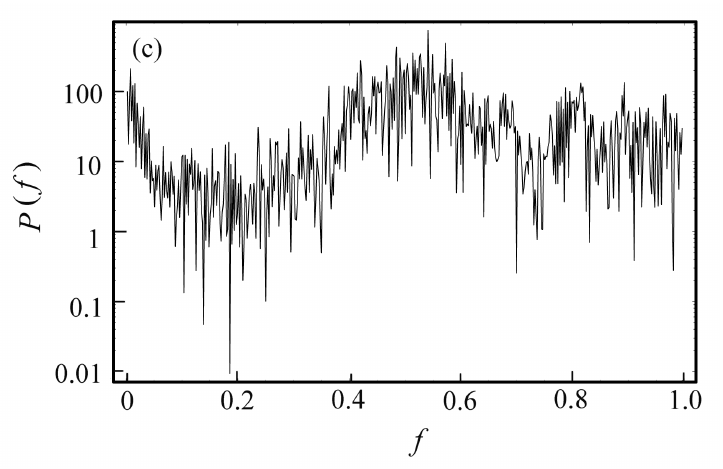}~~~~
\includegraphics[width=66mm]{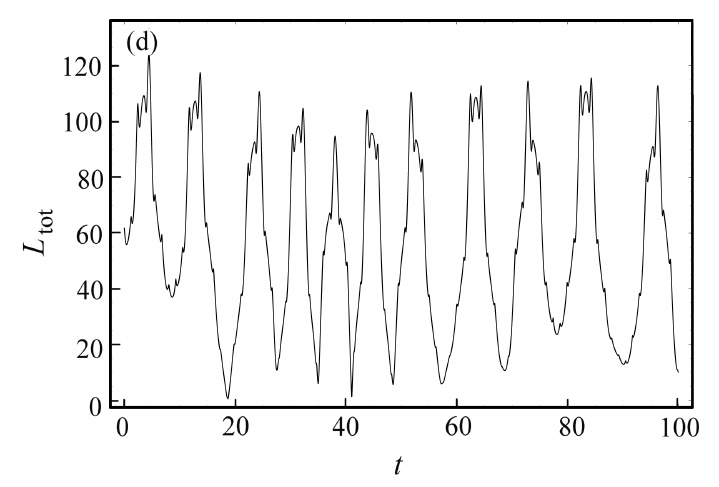}

\caption{\baselineskip 3.6mm (a)
An orbit in the 3D potential, (b) 
The corresponding LCE, (c) 
The $P(f)$ indicator and (d) 
The $L_{\rm tot}$ indicator. The motion {is} chaotic. See {the
}text for details.}
\end{figure}
\begin{figure}

\centering

\includegraphics[width=55mm]{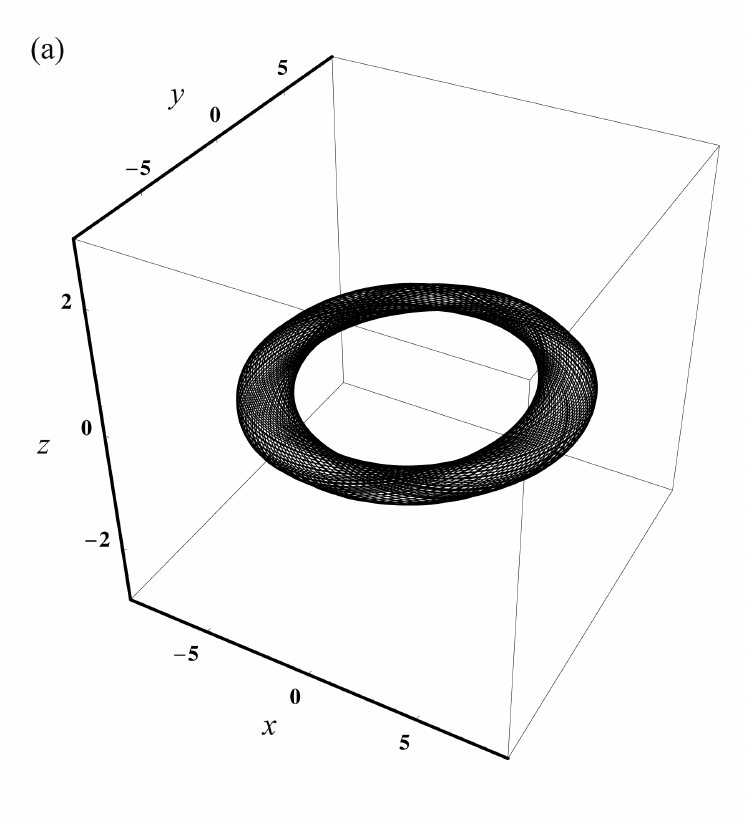}~~~~~
\includegraphics[width=66mm]{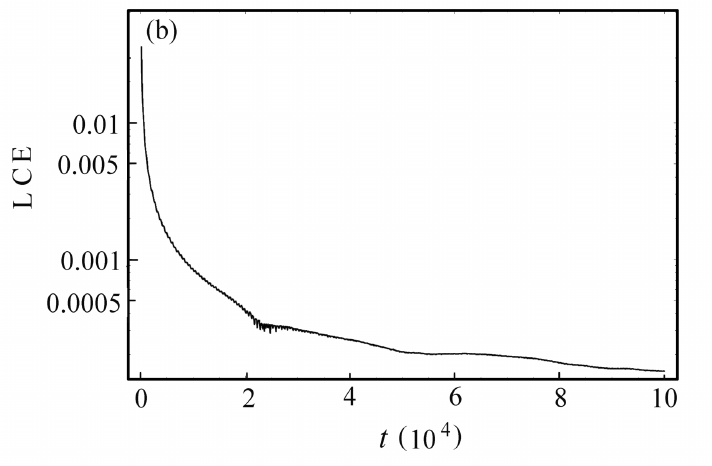}

\vs\vs
\includegraphics[width=65mm]{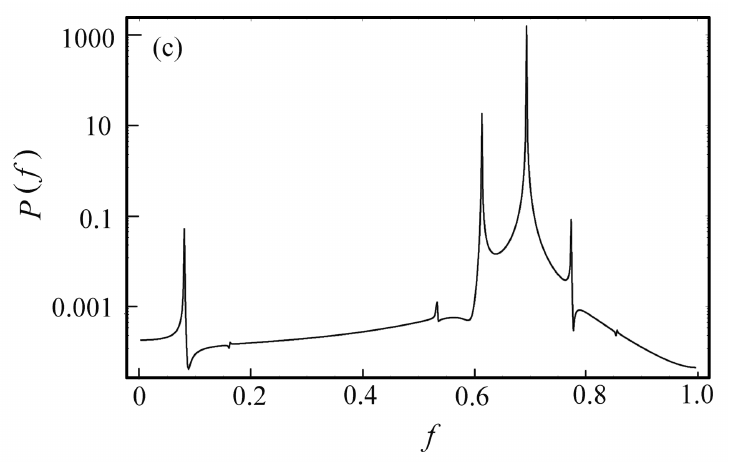}~~~~~~
\includegraphics[width=65mm]{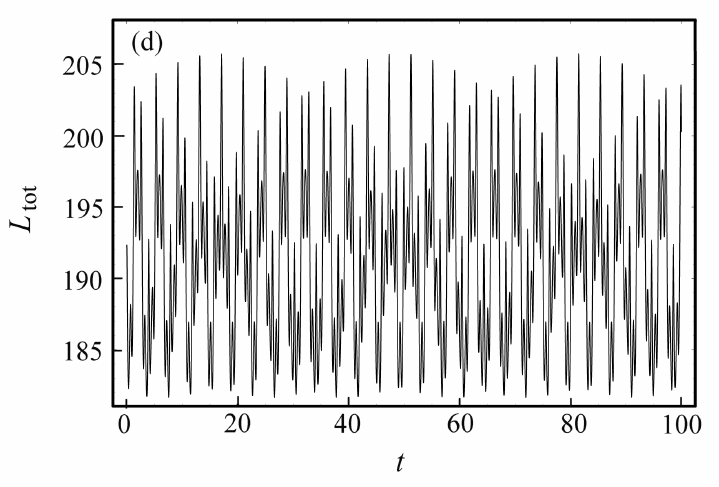}

\begin{minipage}[]{80mm}

\caption{ Similar {to} Fig.~10(a)--(d). The motion is
regular.}\end{minipage}

\vs \vs \centering

\vs
\includegraphics[width=58mm]{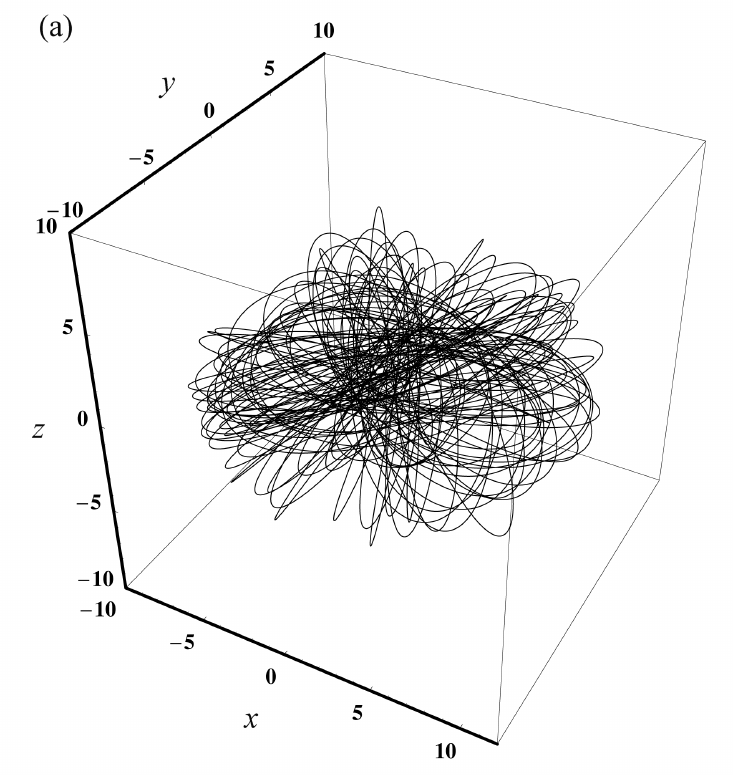}~~~~~
\includegraphics[width=66mm]{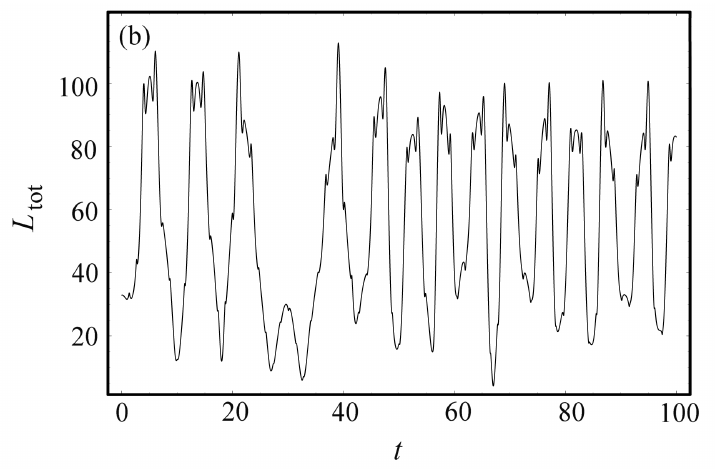}

\begin{minipage}[]{120mm}

\caption{\baselineskip 3.6mm  (a) 
A 3D chaotic orbit and (b) 
 The corresponding
$L_{\rm tot}$. See text for details.}\end{minipage}\vs\vs\vs\vs
\end{figure}

\begin{figure}[h!!]

\vs\centering
\includegraphics[width=55mm]{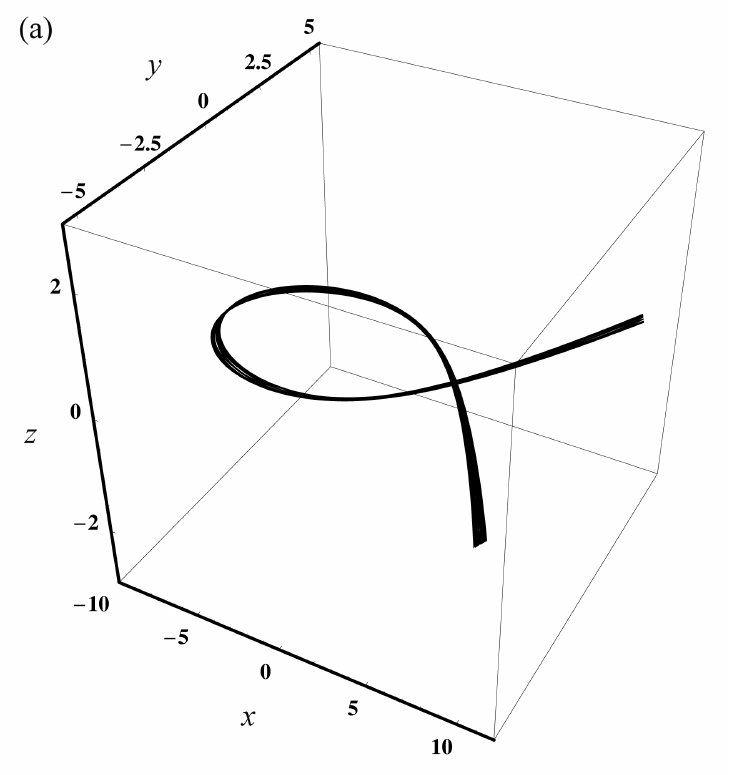}~~~~~
\includegraphics[width=68mm]{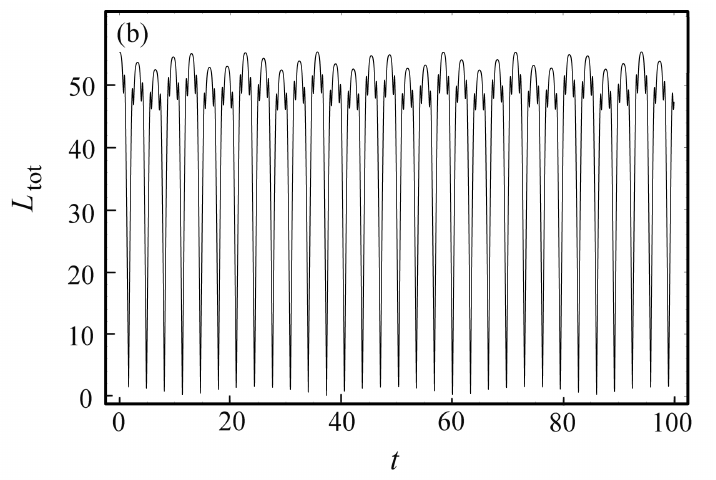}

\begin{minipage}[]{90mm}


\caption{ Similar to Fig.~12(a)--(b) for a quasi-periodic
orbit.}\end{minipage}

\end{figure}

Two more examples of 3D orbits are given in Figure~12(a) and (b)
and Figure~13(a) and (b). Here we present only the orbit and the
$L_{\rm tot}$. In Figure~12(a) and (b) the orbit has initial
conditions $x_0=-1.0$, $p_{x0}=0$ and $z_0=0.5$. The values of all
other parameters and energy $h_3$ are {shown} in Figure~2(c). The
motion is chaotic. In Figures~13(a) and (b) the orbit has initial
conditions $x_0=-9.55$, $p_{x0}=0$ and $z_0=0.1$. The values of
all other parameters and energy $h_3$ are as{ shown} in
Figure~2(d). The motion is regular.

The main conclusion for the study of the 3D model is that
the $L_{\rm tot}$ indicator can give reliable and very fast
results for the character of the orbits. There is no doubt
that the $L_{\rm tot}$ is much faster than the two
other indicators used in this research. Therefore, we can
say that this indicator is a very useful tool for a quick study of
the character of orbits in galactic potentials.



\section{Discussion and conclusions}

In this article, we have studied the regular {and} chaotic character
of motion in a 3D galactic potential. The potential describes the
motion in a triaxial elliptical galaxy with a small asymmetry
surrounded by a dark halo component. Dark halos may have a
variety of shapes (see \citealt{Ioka2000, Olling2000,
Oppenheimer2001, McLin2002, Penton2002, Steidel2002, Wechsler2002,
Papadopoulos2006, Caranicolas2010}). In this investigation we have
used a spherical dark halo component.

In order to distinguish between regular and chaotic motion, we
have introduced and used a new fast indicator{,} the $L_{\rm
tot}$. The validity of the results given by the new indicator was
checked using the LCE and an earlier method used by
\cite{Karanis2007}. We started from the 2D system and then
extended the results to the 3D potential.

The main conclusions of this research are the following:

\begin{itemize}
    \item[(1)] The percentage of chaotic orbits decreases as the mass
of the spherical halo increases. Therefore, the mass of the halo
can be considered as an important physical quantity, acting as a
controller of chaos in galaxies showing small asymmetries.

\item[(2)] We expect to observe a smaller fraction of chaotic
orbits in asymmetric triaxial galaxies with a dense spherical
halo, while the fraction of chaotic orbits would increase in
asymmetric triaxial galaxies surrounded by less dense spherical
halo components.

\item[(3)] It was found that the LCE in both the 2D and the 3D
models decreases as the mass of {the }halo increases, while the
LCE increases as the scale length $c_h$ of the halo increases.
This means that not only the percentage of chaotic orbits, but
also the degree of chaos is affected by the mass or the scale
length of the spherical halo component.

\item[(4)] The $L_{\rm tot}$ gives fast and reliable results
regarding the nature of motion, both in 2D and 3D galactic
potentials. For all calculated orbits{,} the results given by the
$L_{\rm tot}$ coincide with the outcomes obtained using the LCE or
the $P(f)$ method used by \cite{Karanis2007}. The advantage of the
$L_{\rm tot}$ is that{ it} is faster than the above two methods.

\end{itemize}

\begin{acknowledgements}
Our thanks go to an anonymous referee for his useful suggestions and comments.
\end{acknowledgements}


\label{lastpage}

\end{document}